# *Ill-defined Topological Phases in Dispersive Photonic Crystals*

**Filipa R. Prudêncio[1,2*], and Mário G. Silveirinha[1]**

[1]University of Lisbon – Instituto Superior Técnico and Instituto de Telecomunicações, Avenida Rovisco Pais 1, 1049-001 Lisbon, Portugal

[2]Instituto Universitário de Lisboa (ISCTE-IUL), Avenida das Forças Armadas 376, 1600-077 Lisbon, Portugal

## Abstract

In recent years there has been a great interest in topological materials and in their fascinating properties. Topological band theory was initially developed for condensed matter systems, but it can be readily applied to arbitrary wave platforms with little modifications. Thus, the topological classification of optical systems is usually regarded as being mathematically equivalent to that of condensed matter systems. Surprisingly, here we find that both the particle-hole symmetry and the dispersive nature of nonreciprocal photonic materials may lead to situations where the usual topological methods break-down and the Chern topology becomes ill-defined. It is shown that due to the divergence of the density of photonic states in plasmonic systems the gap Chern numbers can be non-integer notwithstanding that the relevant parametric space is compact. In order that the topology of a dispersive photonic crystal is well defined, it is essential to take into account the nonlocal effects in the bulk-materials. We propose two different regularization methods to fix the encountered problems. Our results highlight that the regularized topologies may depend critically on the response of the bulk materials for large **k**.

---

[*] Corresponding author: filipa.prudencio@lx.it.pt

# Main Text

Topological concepts have created exciting opportunities and unveiled hidden connections between different branches of physics, ranging from condensed matter to photonics [1]-[12]. In particular, in the case of optics, topological ideas have offered a more profound understanding of the wave propagation in nonreciprocal platforms, through the link between the topological charge of a medium and the emergence of edge states at the boundaries [13-16].

While the topological band theory was initially developed for condensed matter systems, the theory can be extended with little modifications to nearly arbitrary wave platforms, independent of the nature (fermionic or bosonic) of the system. In fact, the calculation of topological invariants in physical systems can very often be reduced to the problem of characterizing the topology of a two-parameter family of differential (possibly non-Hermitian) linear operators [17]. Apart from a few technical aspects related to the material dispersion [6, 7, 18], the topological band theory of photonic systems is usually regarded as being essentially equivalent to its condensed matter counterpart, with the main differences arising from the physical manifestations of the topology in fluctuation-induced phenomena (e.g., [16, 19, 20]).

The objective of this Letter is to highlight that the material dispersion and the particle-hole symmetry specific of bosonic systems may lead to rather peculiar situations where the standard topological methods breakdown and the Chern-topology becomes ill-defined. We find that in order to ensure that the topology of a dispersive photonic crystal is well-defined it is absolutely essential to take into account the spatially-dispersive response (i.e., dependence of the permittivity on the wave vector) of the bulk materials. In particular, our results demonstrate that rather surprisingly the cut-off associated with the spatial periodicity of photonic crystals is insufficient to guarantee a well-defined topology.

In order to illustrate the ideas, we consider a 2D photonic crystal formed by a hexagonal array of air rods embedded in an electric gyrotropic host (Fig. 1). The gyrotropic material is described by a permittivity tensor of the form $\bar{\varepsilon}_{\text{loc}} = \varepsilon_t \mathbf{1}_t + i\varepsilon_g \hat{\mathbf{z}} \times \mathbf{1}_t + \varepsilon_a \hat{\mathbf{z}} \otimes \hat{\mathbf{z}}$, with $\varepsilon_t = 1 - \frac{\omega_p^2}{\omega^2 - \omega_c^2}$, $\varepsilon_g = \frac{1}{\omega}\frac{\omega_p^2 \omega_c}{\omega_c^2 - \omega^2}$, and $\varepsilon_a = 1 - \frac{\omega_p^2}{\omega^2}$. Here, $\omega_c = -qB_0/m$ is the cyclotron frequency, $q = -e$ is the charge of electrons, $m$ is the effective mass, $\mathbf{B}_0 = B_0 \hat{\mathbf{z}}$ is the bias magnetic field and $\omega_p$ is the plasma frequency. The operators $\times$ and $\otimes$ represent the cross and the tensor products, respectively, and $\mathbf{1}_t = \hat{\mathbf{x}} \otimes \hat{\mathbf{x}} + \hat{\mathbf{y}} \otimes \hat{\mathbf{y}}$. The sign of $\omega_c$ depends on the orientation of the bias magnetic field $B_0$. The magnetic response is assumed trivial ($\mu = \mu_0$). Similar material responses occur naturally in magnetically biased semiconductors, e.g., InSb [21, 22].

It is well-known that the continuous translational symmetry of the bulk medium causes its topology to be ill-defined [18]. It is common understanding that the creation of a crystalline structure, e.g., with a periodic array of inclusions in a host medium, effectively regularizes the topology of the system because it leads to a compact parametric space. Note that in the continuous case the parametric space is the Euclidean plane, whereas in the periodic case it is a Brillouin zone [18]. The parametric space is compact only in the latter case. It is worth mentioning that the calculation of topological invariants of photonic crystals is a rather formidable problem from a computational point of view, and to our best knowledge the characterization of the topology of dispersive photonic crystals was not reported in the literature. Up to now, only a few works studied the topology of photonic crystals using first principles methods [8, 23, 24], but the material dispersion was always ignored.

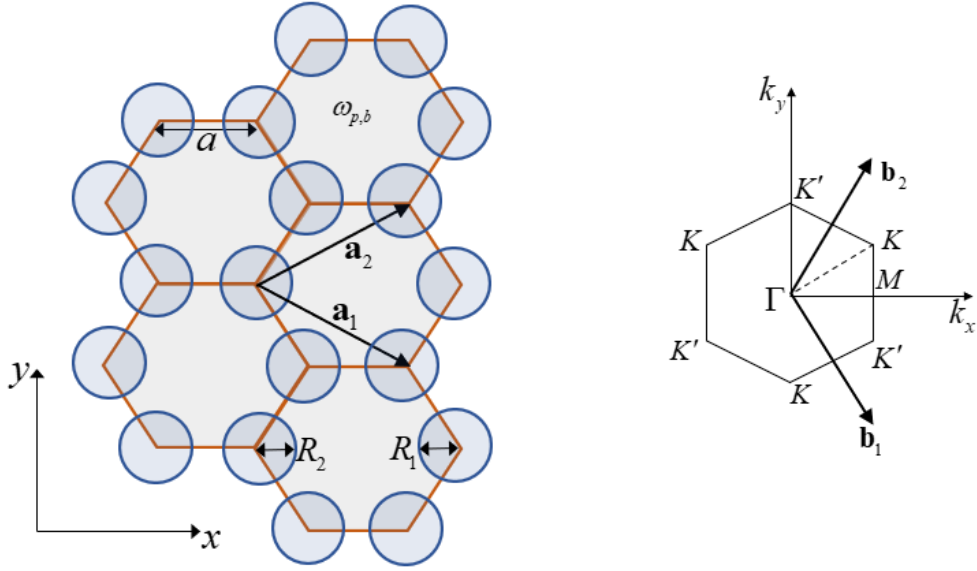

**Fig. 1** Left: Hexagonal array of air rods (blue regions) embedded in an electric gyrotropic host material with plasma frequency $\omega_{p,b}$. Each unit cell contains two air rods (honeycomb lattice) with radius $R_1$ and $R_2$. The distance between nearest neighbors is $a$. The direct lattice primitive vectors are $\mathbf{a}_1$ and $\mathbf{a}_2$, and the reciprocal lattice primitive vectors are denoted by $\mathbf{b}_1$ and $\mathbf{b}_2$. Right: Brillouin zone showing the high symmetry points $\Gamma, M$ and $K$.

Figure 2ai) shows the band structure of a representative dispersive photonic crystal geometry, showing both positive and negative frequencies. The band structure is numerically computed using the plane wave method [25]. To this end, the spectral problem must be first formulated as a standard eigenvalue problem of the type $\hat{L}_\mathbf{k} \cdot \mathbf{Q}_{n\mathbf{k}} = \mathcal{E}_{n\mathbf{k}} \mathbf{Q}_{n\mathbf{k}}$, with $\mathcal{E}_{n\mathbf{k}} = \omega_{n\mathbf{k}}/c$ and $\hat{L}_\mathbf{k}$ a differential operator independent of the frequency. Typically, this entails modeling the effects of the material dispersion with additional variables that represent the internal degrees of freedom of the medium responsible for the dispersive response [18, 26, 27]. Thus, the relevant state vector $\mathbf{Q}_{n\mathbf{k}}$ is formed not only by the electromagnetic fields but also by the aforementioned additional variables (in the present case the current density and charge density in the magnetized plasma). The details are presented in the supplemental material [28].

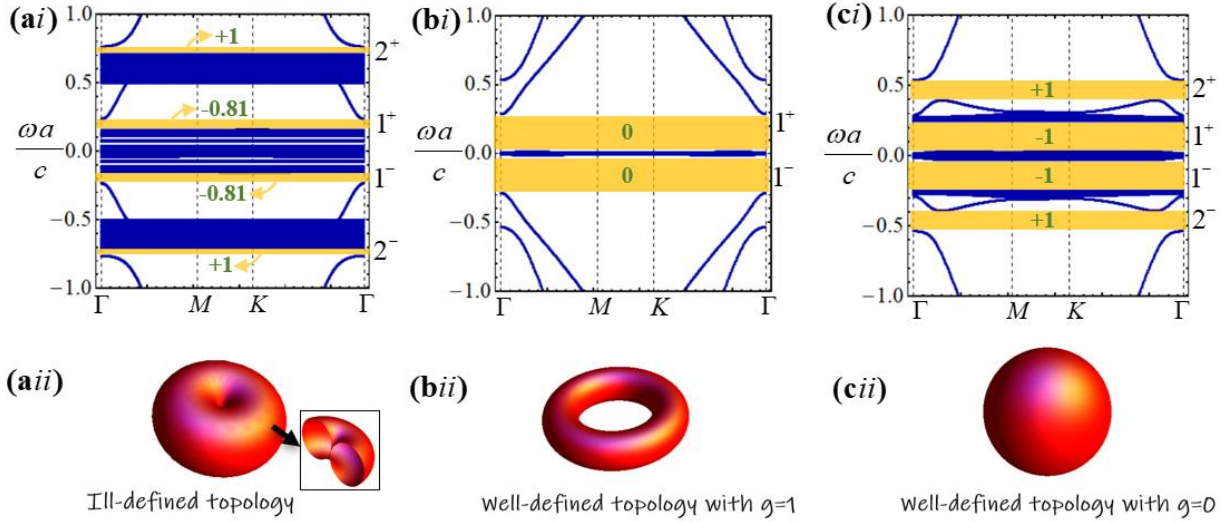

**Fig. 2** Top row: Photonic band structures of dispersive photonic crystals formed by air rods embedded in a magnetized electric plasma. The radius of the rods is $R_1 = R_2 = 0.4a$. The nearest neighbors distance is $a = 0.5c/\omega_p$. Each scalar component of the state vector is expanded with 49 plane waves. The green insets give the numerically calculated gap Chern numbers. The dispersive host material is characterized by **(a)** local model with $\omega_c = \omega_p$, **(b)** hydrodynamic model with $\omega_c = 0.5\omega_p$ and $\beta = 0.5c$, and **(c)** full cutoff model with $\omega_c = 0.5\omega_p$ and $k_{max} = 2\omega_p/c$. Bottom row: Geometrical illustration of the concepts of ill-defined topology and topology regularization. A torus with a vanishing inner radius (panel *a*) has an ill-defined topology because the surface is not differentiable at the central point where the adjoining top and bottom sections touch at a single point (cusp). By either opening a hole in the central region (panel *b* showing a torus) or by separating the top and bottom sections of the surface (panel *c* showing a sphere) one can regularize the topology of the original object. Similar to the electric gyrotropic plasma, the topology of the regularized object depends on the regularization procedure.

Due to the reality of the electromagnetic field (bosonic field), the spectrum is constrained by the particle-hole symmetry $\omega(\mathbf{k}) = -\omega(-\mathbf{k})$, which implies that the positive frequency and negative frequency spectra are linked by a mirror symmetry, consistent with Fig. 2ai). As seen,

5additional variables (in the present case the current density and charge density in the magnetized plasma). The details are presented in the supplemental material [28].

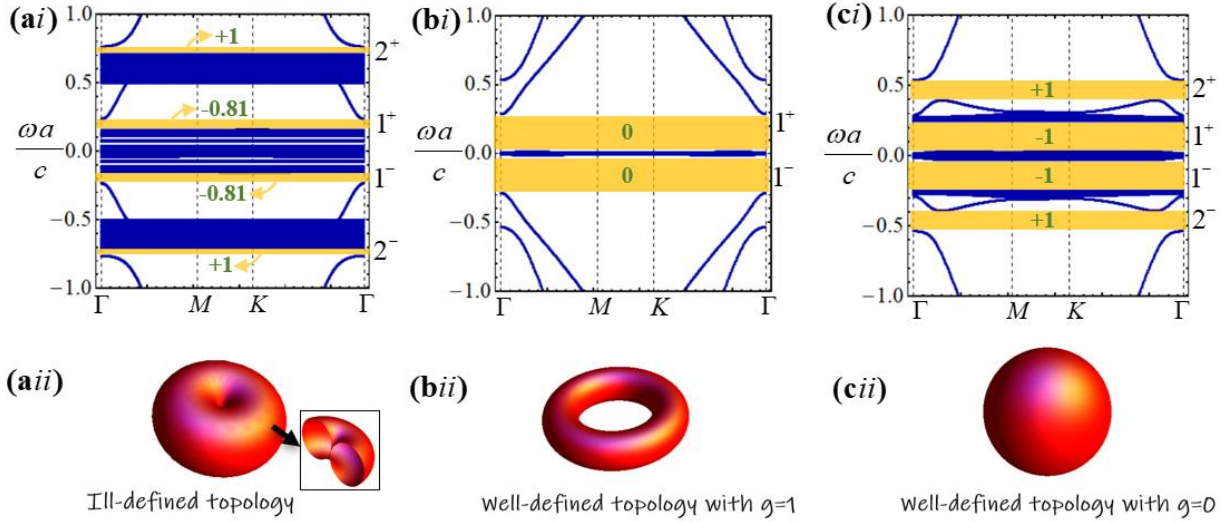

**Fig. 2** Top row: Photonic band structures of dispersive photonic crystals formed by air rods embedded in a magnetized electric plasma. The radius of the rods is $R_1 = R_2 = 0.4a$. The nearest neighbors distance is $a = 0.5c/\omega_p$. Each scalar component of the state vector is expanded with 49 plane waves. The green insets give the numerically calculated gap Chern numbers. The dispersive host material is characterized by **(a)** local model with $\omega_c = \omega_p$, **(b)** hydrodynamic model with $\omega_c = 0.5\omega_p$ and $\beta = 0.5c$, and **(c)** full cutoff model with $\omega_c = 0.5\omega_p$ and $k_{max} = 2\omega_p/c$. Bottom row: Geometrical illustration of the concepts of ill-defined topology and topology regularization. A torus with a vanishing inner radius (panel *a*) has an ill-defined topology because the surface is not differentiable at the central point where the adjoining top and bottom sections touch at a single point (cusp). By either opening a hole in the central region (panel *b* showing a torus) or by separating the top and bottom sections of the surface (panel *c* showing a sphere) one can regularize the topology of the original object. Similar to the electric gyrotropic plasma, the topology of the regularized object depends on the regularization procedure.

Due to the reality of the electromagnetic field (bosonic field), the spectrum is constrained by the particle-hole symmetry $\omega(\mathbf{k}) = -\omega(-\mathbf{k})$, which implies that the positive frequency and negative frequency spectra are linked by a mirror symmetry, consistent with Fig. 2ai). As seen, there are two complete gaps with positive frequency, and evidently two complete gaps with negative frequency. Furthermore, due to the dispersive nature of the material response, one sees an accumulation of an infinite number of branches for frequencies on the order of the bulk plasma frequency $\omega_p$. A similar effect has been reported previously for the case of reciprocal metallic photonic crystals [29, 30], and is rooted in the plasmonic response of a metal which leads to localized resonances that hybridize to form quasi-flat bands.

The topology of the photonic crystal is characterized with the system Green's function $\mathcal{G}_{\mathbf{k}}(\omega) = i(\hat{L}_{\mathbf{k}} - \mathbf{1}\omega)^{-1}$ [17, 24, 31, 32]. The operator $\hat{L}_{\mathbf{k}}$ is parameterized by the real wave vector $\mathbf{k} = k_x \hat{\mathbf{x}} + k_y \hat{\mathbf{y}}$ that determines the Bloch-type boundary conditions in a unit-cell. The poles of the Green's function coincide with the eigenfrequencies $\omega = \omega_{n\mathbf{k}}$ of $\hat{L}_{\mathbf{k}}$. The eigenfrequencies are separated in the complex plane by vertical strips that determine the bandgaps (in the lossless case considered here the eigenfrequencies lie in the real-axis; in the non-Hermitian case they may populate other parts of the complex plane, $\omega = \omega' + i\omega''$) [17, 24]. The gap Chern number of each spectral bandgap can be expressed in terms of the Green's function through an integral in the complex plane over a line parallel to the imaginary axis contained in the bandgap ($\omega = \omega_{\text{gap}} + i\omega''$ with $-\infty < \omega'' < \infty$):

$$\mathcal{C}_{\text{gap}} = \frac{i}{(2\pi)^2} \iint_{B.Z.} d^2\mathbf{k} \int_{\mathcal{E}_{\text{gap}}-i\infty}^{\mathcal{E}_{\text{gap}}+i\infty} d\mathcal{E} \, \text{Tr}\left\{\partial_1 \hat{L}_{\mathbf{k}} \cdot \mathcal{G}_{\mathbf{k}} \cdot \partial_2 \hat{L}_{\mathbf{k}} \cdot \mathcal{G}_{\mathbf{k}}^2\right\}. \tag{1}$$

Here, $\text{Tr}\{...\}$ is the trace operator, $\partial_j = \partial/\partial k_j$ ($j$=1,2) with $k_1 = k_x$ and $k_2 = k_y$, $\mathcal{E}_{\text{gap}} = \omega_{\text{gap}}/c$ is some normalized (real-valued) frequency in the gap [17, 24, 31], and the integral in $\mathbf{k}$ is over the first Brillouin zone (B.Z.). The relevant differential operators and the Green's function are represented by infinite matrices in a plane wave basis [28]. In practice, the relevant integrals and the trace are evaluated through a discretization of the computational domain and a truncation of the plane wave basis. The topological classification with the Green's function does not require knowledge of the eigenfunctions (Bloch modes) of the photonic crystal [18, 31].

The Green's function method determines directly the gap Chern number, i.e., the sum of all Chern numbers below the gap. In our understanding, this is only method that can be applied to compute the topological charge of a dispersive photonic crystal. This is so because typically there are an infinite number of bands below the gap and thereby it is impracticable to apply the standard topological band theory, as that would require evaluating separately the individual contributions of all bands below the gap. The situation is particularly acute in the dispersive case due to the accumulation of an infinite number of branches at finite frequencies, as illustrated in Fig. 2ai).

In the supplemental materials, we present a detailed numerical study of the convergence of the gap Chern numbers of the photonic crystal [28]. Puzzlingly, we find that the low-frequency gap Chern number of the dispersive photonic crystal is not an integer: the converged numerical calculations yield $\mathcal{C}_{\text{gap}}^{1+} \approx -0.81$. On the other hand, for the positive high-frequency gap the converged Chern number is $\mathcal{C}_{\text{gap}}^{2+} = 1$, which as expected is an integer and is insensitive to perturbations of the geometry that do not close the gap. The ill-defined topology of the low-frequency gap seemingly contradicts the Chern theorem, which naively is expected to apply because the parametric space (B.Z.) is a compact set with no boundary. In the following, we argue that the topology may be ill-defined due to two reasons: i) the existence of an infinite number of bands below the gap, and ii) the accumulation of an infinite number of branches at a single frequency (Fig. 3). The latter reason is the relevant one for the system under analysis.

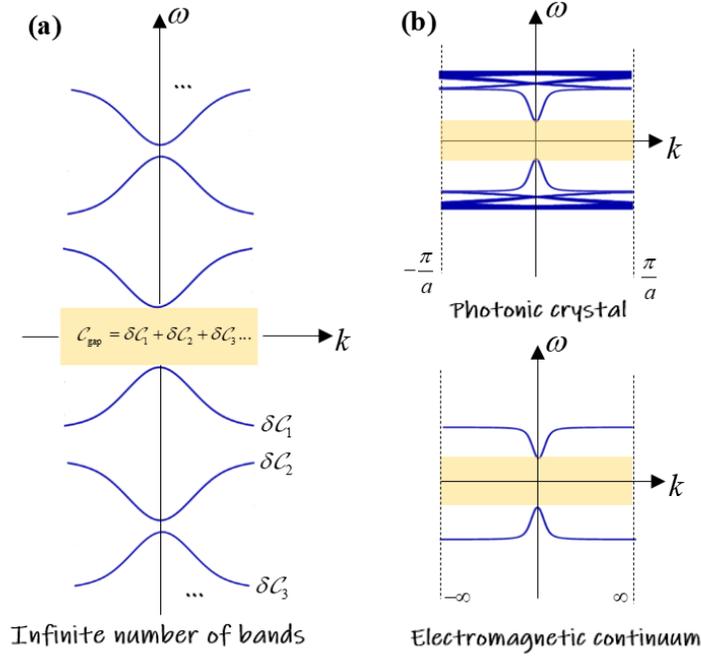

**Fig. 3 (a)** Illustration of the band structure of a photonic crystal. Due to the particle-hole symmetry the gap Chern number is generally given by an infinite sum of integers, corresponding to the sum of the Chern numbers of all the bands below the gap. **(b)** The accumulation of branches at a single frequency in a photonic crystal (top panel) may be regarded as a consequence of the band folding of the dispersion of the bulk-material (bottom panel).

As previously noted, photonic systems are constrained by the particle-hole symmetry. Then, different from electronic systems which have a well-defined "ground", in the photonic case it is possible to have an infinite number of bands below a bandgap. Indeed, the number of bands with negative-frequency is typically infinite. There is a widespread belief that the total topological charge of negative frequency bands vanishes, and so they do not play any role. Such an understanding is flawed: for a counterexample in an electromagnetic continuum see Ref. [16]. Below we also present a counterexample in a photonic crystal. In fact, the particle-hole symmetry only implies that the total topological charge of positive frequency bands has the opposite sign of the charge of negative frequency bands, nothing more.

The potential problem of having an infinite number of bands below the gap is illustrated in Fig. 3a. Consider the bandgap shaded in yellow. As the corresponding gap Chern number is given by $\mathcal{C}_{gap} = \delta\mathcal{C}_1 + \delta\mathcal{C}_2 + ...$, with $\delta\mathcal{C}_i$ the Chern numbers of the bands below the gap, it follows that $\mathcal{C}_{gap}$ is given by an infinite sum of integers. In general, such a series is divergent. For example, suppose that $\delta\mathcal{C}_i = (-1)^i$. Then, the gap Chern number is given by $\mathcal{C}_{gap} = -1+1-1+...$, which is evidently not convergent in the usual sense. In fact, a series of the type $\mathcal{C}_{gap} = \delta\mathcal{C}_1 + \delta\mathcal{C}_2 + ...$ with $\delta\mathcal{C}_i$ integer can converge only when the number of non-vanishing terms is finite! Fortunately, there is a relatively simple way to enforce such a constraint for the scenario of Fig. 3a. Intuitively, provided the response of the nonreciprocal materials approaches that of the vacuum ($\bar{\varepsilon} \to \varepsilon_0 \mathbf{1}$) when $\omega \to \infty$, the topological charge of bands with large frequency is expected to be trivial (i.e., $\delta\mathcal{C}_i = 0$) for a sufficiently large $i$. Note that the constraint $\bar{\varepsilon}_{\omega \to \infty} \to \varepsilon_0 \mathbf{1}$ is a consequence of the Kramers-Kronig formulas [33], and is satisfied by the dispersive model considered here.

Interestingly, a related problem may occur even when the physical constraint $\bar{\varepsilon}_{\omega \to \infty} \to \varepsilon_0 \mathbf{1}$ is enforced. In fact, suppose that a certain band is formed by an infinite number of "branches", i.e., that there is an accumulation of eigenvalues at a finite frequency (see Fig. 3b, top panel). Intuitively, the Chern number $\delta\mathcal{C}$ of the band is a sum of an infinite number of integer contributions, resulting in an ill-defined topology for the same reasons as before. A more rigorous argument is developed in the supplementary materials [28]. The described situation is precisely what happens in Fig. 2ai): the topological charge of the band in between gaps $1^+$ and $2^+$ is given by $\delta\mathcal{C} = \mathcal{C}_{gap}^{2+} - \mathcal{C}_{gap}^{1+} = 1.81$. In contrast, the topological charge of all the bands above gap $2^+$ is $\delta \mathcal{C} = 0 - \mathcal{C}_{\text{gap}}^{2+} = -1$; the topology of these bands is well-defined because above gap $2^+$ there are no plasmonic-type localized resonances, and thereby there is no accumulation of eigenvalues at a finite frequency.

The problem identified in the previous paragraph can be fixed in two ways. The first solution is to guarantee that there is no accumulation of branches at a finite frequency. For the particular system under study, this can be done taking into account the effects of charge diffusion in the electron gas using the so-called "hydrodynamic" model [22, 34-36]. The strength of the diffusion term is determined by the velocity $\beta$, which typically corresponds to the velocity of electrons at the Fermi level. The physical origin of the diffusion term is the electron-electron repulsive interactions. Figure 2bi) shows the photonic band structure of a dispersive photonic crystal, with the diffusion effects modeled by $\beta = 0.5c$ (this unrealistically large value of $\beta$ is chosen to have larger gaps and speed up the numerical calculations; the topology of the gap $1^+$ is independent of the value of $\beta > 0$). As seen, a nonzero value of $\beta$ may change considerably the band structure as compared to the local model of Figure 2ai). Now, there is a single positive-frequency bandgap determined by $0.01 < \omega c/a < 0.29$. This bandgap remains open if $\beta$ is decreased continuously down to zero (see [28]), and hence it can be identified with the gap $1^+$ of the local model. As seen in Fig. 2bi), the diffusion effects prevent the accumulation of bands at a single frequency. In agreement with this property, we find that the gap Chern number is now an integer, $\mathcal{C}_{\text{gap}}^{1+} = 0$, and thereby the dispersive photonic crystal has a trivial topology. The convergence analysis is reported in the supplemental material [28].

Next, we discuss a second and more general solution to regularize the topology of the dispersive photonic crystal. To begin with, we note that the accumulation of branches at a single frequency can be regarded as a consequence of the band-folding of the dispersion of the bulk-host material (see Fig. 3b). From this point of view, the ill-defined topology of the photonic crystal is inherited from the ill-defined topology of the host material [18]. This suggests that a regularization of the topology of the host material may also fix the topology of the photonic crystal. Reference [18] introduced a general solution to regularize the topology of an electromagnetic continuum; the procedure is based on the introduction of a full spatial cutoff $k_{max}$ that guarantees that $\bar{\varepsilon}_{\mathbf{k}\to\infty} \to \varepsilon_0 \mathbf{1}$. In other words, the material response is suppressed for large wave vectors so that it becomes identical to that of free-space. This can be implemented by modifying the original local response ( $\bar{\varepsilon}_{loc}$ ) in such a way that

$$\bar{\varepsilon}_{nonloc} = \varepsilon_0 \mathbf{1} + \frac{1}{1+k^2/k_{max}^2}(\bar{\varepsilon}_{loc} - \varepsilon_0 \mathbf{1})$$ [18]. Note that for values of $k \ll k_{max}$ the cutoff leaves the original response almost unchanged. It is relevant to underline that the constraint $\bar{\varepsilon}_{\mathbf{k}\to\infty} \to \varepsilon_0 \mathbf{1}$ is rather physical, as it is a necessary consequence of the Kramers-Kronig relations for media with spatial dispersion and of the Riemann-Lebesgue lemma [33, 37]. Figure 2ci) reports the band structure of a dispersive photonic crystal with the spatial cut-off $k_{max} = 2\omega_p/c$. The relevant formalism and the details of the numerical implementation are given in [28]. Different from the hydrodynamic model, the two original bandgaps are now preserved as they remain open when the cutoff is introduced (see [28]). Moreover, different from the hydrodynamic model, there is still an accumulation of branches at a single frequency. However, the numerical calculations reveal that the topology of the two bandgaps is well defined, yielding $C_{gap}^{2+} = 1$ and $C_{gap}^{1+} = -1$ (see the supplemental material [28] for the convergence analysis and for the explanation why the

cutoff $k_{max}$ regularizes the topology). In particular, this example nicely illustrates that the total topological charge of the negative frequency bands can be nontrivial as $\mathcal{C}_{gap}^{1-} \neq 0$.

Remarkably, even though both the hydrodynamic and the full cutoff models regularize the topology of the gap $1^+$ of the original problem, they lead to a different gap Chern number. This property can be explained in a geometrical way (see the bottom row of Fig. 2). The original (local) dispersive photonic crystal may be regarded as the counterpart of a non-differentiable geometric surface, e.g., a torus with vanishing radius as shown in Fig. 2ai). As the topology of a geometric surface is determined by the number of holes (genus), the topology of a torus with vanishing inner radius is ill-defined because the top and bottom sections touch at exactly one point. The surface topology can be regularized with a negligibly weak perturbation of the original shape. One option is to insert a hole in the middle region to create a torus with a nonzero inner radius (Fig. 2bii). Another option is to separate the top and bottom sections to obtain an object topologically equivalent to a sphere (Fig. 2cii). Evidently, the topology of the two regularized objects (the genus) is different. In fact, it depends on the regularization procedure, exactly in the same manner as the topology of the regularized photonic crystal depends if one adopts the hydrodynamic model or the full cutoff model.

In summary, it was shown that surprisingly the periodicity of a dispersive photonic crystal does not guarantee a well-defined topology. It was highlighted that both the particle-hole symmetry and the accumulation of resonances at a single frequency (e.g., divergence of the density of photonic states due to plasmonic resonances and band folding) can lead to ill-defined topologies, where the gap Chern number is not an integer. We proposed two simple solutions to fix the encountered problem. The first solution exploits charge diffusion to prevent the accumulation of bands at a single frequency; the second solution, which is applicable in any scenario, ensures that the contribution of most branches is trivial by suppressing the material response for large wave-vectors. An important corollary of our findings is that the topology of any dispersive photonic crystal generally depends critically on the high-spatial frequency response of the involved bulk materials. To conclude, we observe that nonreciprocal systems with ill-defined topologies are interesting on their own [38-42], as they may enable the emergence of topological energy sinks that can be useful for energy harvesting [38].

### Acknowledgements

This work is supported in part by the IET under the A F Harvey Engineering Research Prize, by the Simons Foundation under the award 733700 (Simons Collaboration in Mathematics and Physics, "Harnessing Universal Symmetry Concepts for Extreme Wave Phenomena"), and by Fundação para a Ciência e a Tecnologia and Instituto de Telecomunicações under project UID/EEA/50008/2020.

# Supplemental Information:

## "Ill-defined Topological Phases in Dispersive Photonic Crystals"

**Filipa R. Prudêncio[1,2*], and Mário G. Silveirinha[1]**

[1]University of Lisbon – Instituto Superior Técnico and Instituto de Telecomunicações, Avenida Rovisco Pais 1, 1049-001 Lisbon, Portugal

[2]Instituto Universitário de Lisboa (ISCTE-IUL), Avenida das Forças Armadas 376, 1600-077 Lisbon, Portugal

The supplemental information provides additional details on A) Formulation of the spectral problems as standard eigenvalue problems, B) Representation of the operator $\hat{L}_\mathbf{k}$ in a plane wave basis. C) Continuous evolution of the band diagrams of the hydrodynamic and full cutoff models to the band diagram of the local model. D) Detailed numerical study of the convergence of the gap Chern number. E) The origin of the ill-defined Chern number.

## A. Formulation of the spectral problem as a standard eigenvalue problem

It is convenient to formulate the wave-propagation problem in the form of a Schrödinger-type equation, both for the band structure calculations and for the topological classification. In the following sub-sections, this is done for I) the local and hydrodynamic models, and II) the full cutoff model.

[*] Corresponding author: filipa.prudencio@lx.it.pt

**I) Local and hydrodynamic models**

To begin with, consider the Maxwell's equations in the time domain,

$$-i\nabla \times \mathbf{E} = i\mu_0 \partial_t \mathbf{H}, \qquad i(\nabla \times \mathbf{H} - \mathbf{j}) = i\varepsilon_0 \partial_t \mathbf{E}, \qquad (S1)$$

where $\mathbf{E}$ and $\mathbf{H}$ are the electric and magnetic fields, $\mathbf{D}$ and $\mathbf{B}$ are the electric displacement and the induction fields, $\mathbf{j}$ is the electric current density, and $\varepsilon_0$ and $\mu_0$ are the vacuum permittivity and permeability, respectively. The continuity equation is given by,

$$\partial_t \rho + \nabla \cdot \mathbf{j} = 0, \qquad (S2)$$

where $\rho$ is the charge density. The current and charge densities model the response of the dispersive electric gyrotropic material. The relevant transport equation for a free electron gas biased with a static magnetic field ($\mathbf{B}_0 = B_0 \hat{\mathbf{z}}$) is [S1-S3]:

$$\partial_t \mathbf{j} = \varepsilon_0 \omega_p^2 \mathbf{E} + \frac{q}{m} \mathbf{j} \times \mathbf{B}_0 - \beta^2 \nabla \rho, \qquad (S3)$$

where $\omega_p$ is the plasma frequency, $\beta$ is the diffusion velocity, and $q = -e$ is the charge of electrons. The diffusion coefficient is taken $\beta = 0$ in the local model.

We focus our attention on H-polarized waves with $\mathbf{H} = H_z \hat{\mathbf{z}}$, $\mathbf{E} = E_x \hat{\mathbf{x}} + E_y \hat{\mathbf{y}}$, $\mathbf{j} = j_x \hat{\mathbf{x}} + j_y \hat{\mathbf{y}}$, and $\partial/\partial z = 0$. For this polarization, Eqs. (S1)-(S3) can be rewritten as a Schrödinger-type equation

$$\hat{L} \cdot \mathbf{Q} = \frac{1}{c} i \partial_t \mathbf{Q}, \qquad (S4)$$

with the state vector $\mathbf{Q}$ given by $\mathbf{Q} = \begin{bmatrix} E_x & E_y & \tilde{H}_z & \tilde{j}_x & \tilde{j}_y & \tilde{\rho} \end{bmatrix}^\mathrm{T}$ and "T" the transpose operator. We introduced $\tilde{\mathbf{j}} = \eta_0 \mathbf{j}$, $\tilde{H}_z = \eta_0 H_z$, $\tilde{\rho} = \eta_0 c \rho$ with $\eta_0$ the vacuum wave impedance and $c$ the speed of light. The differential operator $\hat{L} = \hat{L}(-i\nabla, \mathbf{r})$ is given by:

$$\hat{L}(-i\nabla,\mathbf{r}) = \begin{pmatrix} 0 & 0 & i\partial_y & -i & 0 & 0 \\ 0 & 0 & -i\partial_x & 0 & -i & 0 \\ i\partial_y & -i\partial_x & 0 & 0 & 0 & 0 \\ i\omega_p^2/c^2 & 0 & 0 & 0 & -i\omega_c/c & -\beta^2/c^2 i\partial_x \\ 0 & i\omega_p^2/c^2 & 0 & i\omega_c/c & 0 & -\beta^2/c^2 i\partial_y \\ 0 & 0 & 0 & -i\partial_x & -i\partial_y & 0 \end{pmatrix}. \quad (S5)$$

In the above, $\omega_c = -qB_0/m$ is the cyclotron frequency. It is convenient to regard the plasma frequency as periodic function of position, $\omega_p = \omega_p(\mathbf{r})$. This enables us to model the air rods of the photonic crystal (Fig. 1 of the main text) as regions wherein $\omega_p = 0$.

It is relevant to note that (for a homogeneous structure) the operator $\hat{L}$ effectively describes the propagation in an electric gyrotropic medium with the equivalent permittivity [S1]:

$$\frac{1}{\varepsilon_0}\overline{\varepsilon}_{\text{hydro}}(\omega,\mathbf{k}) = \mathbf{1} - \frac{\omega_p^2}{\omega^2}\frac{\Delta}{\Delta+\omega_c^2}\mathbf{1}_t - \frac{\omega_p^2}{\omega^2}\hat{\mathbf{z}}\otimes\hat{\mathbf{z}} + \frac{\omega_p^2}{\omega^2}\frac{\beta^2 \mathbf{k}\otimes\mathbf{k}}{\Delta+\omega_c^2} + \frac{1}{\omega}\frac{i\omega_c\omega_p^2}{\Delta+\omega_c^2}\hat{\mathbf{z}}\times\mathbf{1} \quad (S6)$$

where $\Delta = \beta^2 k^2 - \omega^2$, $k^2 = \mathbf{k}\cdot\mathbf{k}$, $\mathbf{1} = \hat{\mathbf{x}}\otimes\hat{\mathbf{x}} + \hat{\mathbf{y}}\otimes\hat{\mathbf{y}} + \hat{\mathbf{z}}\otimes\hat{\mathbf{z}}$ is the identity matrix and $\mathbf{1}_t = \hat{\mathbf{x}}\otimes\hat{\mathbf{x}} + \hat{\mathbf{y}}\otimes\hat{\mathbf{y}}$. The above formula can be derived by solving Eq. (S3) with respect to $\mathbf{j}$ (as a function of the electric field), assuming a spatial variation of the state vector of the type $e^{i\mathbf{k}\cdot\mathbf{r}}$ with $\mathbf{k} = k_x\hat{\mathbf{x}} + k_y\hat{\mathbf{y}}$ the wave vector. Thus, the frequency-independent operator $\hat{L}$ effectively models the propagation in a dispersive medium.

The Bloch modes of the photonic crystal correspond to solutions of the form $\mathbf{Q} = \mathbf{Q}_{n\mathbf{k}}(\mathbf{r})e^{i\mathbf{k}\cdot\mathbf{r}}$, with $\mathbf{Q}_{n\mathbf{k}}(\mathbf{r})$ the periodic envelope and $\mathbf{k}$ the wave vector. For a photonic crystal with the honeycomb symmetry, the direct lattice primitive vectors are:

$$\mathbf{a}_1 = \frac{a}{2}\left(3\hat{\mathbf{x}} - \sqrt{3}\hat{\mathbf{y}}\right), \qquad \mathbf{a}_2 = \frac{a}{2}\left(3\hat{\mathbf{x}} + \sqrt{3}\hat{\mathbf{y}}\right), \quad (S7)$$

where $a$ is the distance between nearest neighbors. The corresponding reciprocal lattice primitive vectors $\mathbf{b}_1$ and $\mathbf{b}_2$ are given by:

$$\mathbf{b}_1 = \frac{2\pi}{a}\left(\frac{1}{3}\hat{\mathbf{x}} - \frac{1}{\sqrt{3}}\hat{\mathbf{y}}\right), \qquad \mathbf{b}_2 = \frac{2\pi}{a}\left(\frac{1}{3}\hat{\mathbf{x}} + \frac{1}{\sqrt{3}}\hat{\mathbf{y}}\right). \tag{S8}$$

Evidently, the spectral problem reduces to the standard eigenvalue problem:

$$\hat{L}_{\mathbf{k}} \cdot \mathbf{Q}_{n\mathbf{k}} = \frac{\omega_{n\mathbf{k}}}{c}\mathbf{Q}_{n\mathbf{k}}, \qquad \text{with} \qquad \hat{L}_{\mathbf{k}} \equiv \hat{L}(-i\nabla + \mathbf{k}, \mathbf{r}). \tag{S9}$$

### II) Full spatial cutoff model

As discussed in the main text, the topology of the local dispersive photonic crystal can be regularized either by using the hydrodynamic model with a $\beta \neq 0$ (see the previous subsection), or, alternatively, by enforcing a full spatial cutoff that suppresses the material response for large wavevectors [S4-S5]. For a homogeneous medium, the effect of the full cutoff is to modify the original material response ($\bar{\varepsilon}_{\text{loc}}(\omega)$) in such a way that

$$\bar{\varepsilon}_{\text{nonloc}}(\omega, \mathbf{k}) = \varepsilon_0 \mathbf{1} + \frac{1}{1 + k^2/k_{\text{max}}^2}\left(\bar{\varepsilon}_{\text{loc}}(\omega) - \varepsilon_0 \mathbf{1}\right). \tag{S10}$$

In the above, $k_{\text{max}}$ determines the cut-off value. For $k \ll k_{\text{max}}$ the modified response is nearly coincident with $\bar{\varepsilon}_{\text{loc}}(\omega)$, whereas for $k \gg k_{\text{max}}$ the response of the medium is nearly suppressed ($\bar{\varepsilon}_{\text{nonloc}}(\omega, \mathbf{k}) \approx \varepsilon_0 \mathbf{1}$). The full cutoff can be enforced in the system of equations (S1)-(S3) by setting $\beta = 0$ (original local model) and by replacing $\mathbf{j}$ in equation (S1) by $\left(-k_{\text{max}}^{-2}\nabla^2 + 1\right)^{-1}\mathbf{j}$ [S5]. Such a procedure leads to:

$$-i\nabla \times \mathbf{E} = i\mu_0 \partial_t \mathbf{H}, \qquad i\left[\nabla \times \mathbf{H} - \left(-k_{\text{max}}^{-2}\nabla^2 + 1\right)^{-1}\mathbf{j}\right] = i\varepsilon_0 \partial_t \mathbf{E} \tag{S11}$$

$$\partial_t \mathbf{j} = \varepsilon_0 \omega_p^2 \mathbf{E} + \frac{q}{m} \mathbf{j} \times \mathbf{B}_0.  \quad (S12)$$

The charge continuity equation is omitted because it does not play any relevant role when $\beta = 0$. Similar to the previous subsection, the system of equations (S11)-(S12) can be written in the form $\hat{L} \cdot \mathbf{Q} = \frac{1}{c} i \partial_t \mathbf{Q}$, with the state vector now given by $\mathbf{Q} = \begin{bmatrix} E_x & E_y & \tilde{H}_z & \tilde{j}_x & \tilde{j}_y \end{bmatrix}^T$ and the operator $\hat{L}$ defined by:

$$\hat{L} = \begin{pmatrix} 0 & 0 & i\partial_y & -i\left(-k_{max}^{-2}\nabla^2 + 1\right)^{-1} & 0 \\ 0 & 0 & -i\partial_x & 0 & -i\left(-k_{max}^{-2}\nabla^2 + 1\right)^{-1} \\ i\partial_y & -i\partial_x & 0 & 0 & 0 \\ i\omega_p^2/c^2 & 0 & 0 & 0 & -i\omega_c/c \\ 0 & i\omega_p^2/c^2 & 0 & i\omega_c/c & 0 \end{pmatrix}. \quad (S13)$$

As in the previous subsection, $\omega_p = \omega_p(\mathbf{r})$ is regarded as a function of position. The Bloch modes of the photonic crystal are still determined from Eq. (S9), but now with $\hat{L}$ defined as above.

## B. Plane wave representation of the differential operator $\hat{L}_\mathbf{k}$

In this section, we obtain a plane wave representation of the operator $\hat{L}_\mathbf{k} = \hat{L}(-i\nabla + \mathbf{k}, \mathbf{r})$ for the local formulation, for the hydrodynamic model and for the full cutoff model.

### I) Local and hydrodynamic models

The operator $\hat{L}$ is determined by Eq. (S5) for both the local ($\beta = 0$) and hydrodynamic models. Suppose that the envelope of the state vector ($\mathbf{Q}_\mathbf{k} = \mathbf{Q} e^{-i\mathbf{k}\cdot\mathbf{r}}$) is expanded in a plane wave basis as follows:

$$\mathbf{Q_k} = \sum_{\mathbf{J}=(j_1,j_2)} \mathbf{c_J} e^{i\mathbf{G_J}\cdot\mathbf{r}}, \quad \text{with} \quad \mathbf{G_J} = j_1\mathbf{b}_1 + j_2\mathbf{b}_2. \tag{S14}$$

Here, $\mathbf{c_J} = \begin{bmatrix}[E_{x,\mathbf{J}}] & [E_{y,\mathbf{J}}] & [\tilde{H}_{z,\mathbf{J}}] & [\tilde{j}_{x,\mathbf{J}}] & [\tilde{j}_{y,\mathbf{J}}] & [\tilde{\rho}_{\mathbf{J}}]\end{bmatrix}^T$ are the coefficients of the plane wave expansion, $\mathbf{G_J}$ is a generic reciprocal lattice primitive vector, and $\mathbf{J}=(j_1,j_2)$ is a double index of integers. Note that $\hat{L}\cdot\mathbf{Q} = e^{i\mathbf{k}\cdot\mathbf{r}}\hat{L}_\mathbf{k}\cdot\mathbf{Q_k}$. The coordinates of the vector $\hat{L}_\mathbf{k}\cdot\mathbf{Q_k}$ in the considered basis can be formally expressed as $\mathbf{L_k}\cdot[\mathbf{c_J}]$, with $\mathbf{L_k}$ the infinite matrix that represents the operator $\hat{L}_\mathbf{k}$ in the plane-wave basis. Straightforward calculations show that $\mathbf{L_k}$ is given by:

$$\mathbf{L_k} = \begin{pmatrix} 0 & 0 & \left[-(\mathbf{k}+\mathbf{G_I})\cdot\hat{\mathbf{y}}\delta_{\mathbf{I,J}}\right] & \left[-i\delta_{\mathbf{I,J}}\right] & 0 & 0 \\ 0 & 0 & \left[(\mathbf{k}+\mathbf{G_I})\cdot\hat{\mathbf{x}}\delta_{\mathbf{I,J}}\right] & 0 & \left[-i\delta_{\mathbf{I,J}}\right] & 0 \\ \left[-(\mathbf{k}+\mathbf{G_I})\cdot\hat{\mathbf{y}}\delta_{\mathbf{I,J}}\right] & \left[(\mathbf{k}+\mathbf{G_I})\cdot\hat{\mathbf{x}}\delta_{\mathbf{I,J}}\right] & 0 & 0 & 0 & 0 \\ i\frac{1}{c^2}\left[p_{\omega_p^2,\mathbf{I-J}}\right] & 0 & 0 & 0 & \left[-i\frac{\omega_c}{c}\delta_{\mathbf{I,J}}\right] & \frac{\beta^2}{c^2}\left[(\mathbf{k}+\mathbf{G_I})\cdot\hat{\mathbf{x}}\delta_{\mathbf{I,J}}\right] \\ 0 & i\frac{1}{c^2}\left[p_{\omega_p^2,\mathbf{I-J}}\right] & 0 & \left[+i\frac{\omega_c}{c}\delta_{\mathbf{I,J}}\right] & 0 & \frac{\beta^2}{c^2}\left[(\mathbf{k}+\mathbf{G_I})\cdot\hat{\mathbf{y}}\delta_{\mathbf{I,J}}\right] \\ 0 & 0 & 0 & \left[(\mathbf{k}+\mathbf{G_I})\cdot\hat{\mathbf{x}}\delta_{\mathbf{I,J}}\right] & \left[(\mathbf{k}+\mathbf{G_I})\cdot\hat{\mathbf{y}}\delta_{\mathbf{I,J}}\right] & 0 \end{pmatrix}$$
.(S15)

Each of the entries of the matrix is itself an infinite matrix. The double indices $\mathbf{I}=(i_1,i_2)$ and $\mathbf{J}=(j_1,j_2)$ label a generic element $a_{\mathbf{I,J}}$ of a generic submatrix $[a_{\mathbf{I,J}}]$. In the above, $\delta_{\mathbf{I,J}}$ is the Kronecker $\delta$-symbol. Most of the entries of $\mathbf{L_k}$ are represented by diagonal matrices because the unique terms of $\hat{L}_\mathbf{k}$ that depend on the spatial coordinates are the ones associated with $\omega_p = \omega_p(\mathbf{r})$. It should be noted that when $\omega_p = 0$ (air rods regions) the response of the material is the same as in free-space, independent of the value of $\omega_c$ and $\beta$. Finally, $p_{\omega_p^2,\mathbf{I}}$ represents the Fourier coefficients of the periodic function $g(\mathbf{r}) = \omega_p^2(\mathbf{r})$. For the geometry of Fig. 1 of the main text, $g(\mathbf{r}) = g_b$ in the background (host) region of the unit cell, and $g = g_1$ for the rod centered at $\mathbf{r}_{0,1} = -a\hat{\mathbf{x}}$ with radius $R_1$, and $g = g_2$ for the rod centered at $\mathbf{r}_{0,2} = \mathbf{0}$ with radius $R_2$ (for the problem under study, $g_b = \omega_p^2$ and $g_1 = g_2 = 0$). A straightforward analysis shows that [S6, S7]:

$$p_{g,\mathbf{I}} = g_b \delta_{\mathbf{I},0} + \sum_{l=1,2} f_{V,l} \left(g_l - g_b\right) e^{-i\mathbf{G}_\mathbf{I} \cdot \mathbf{r}_{0,l}} \frac{2J_1\left(|\mathbf{G}_\mathbf{I}|R_l\right)}{|\mathbf{G}_\mathbf{I}|R_l}, \tag{S16}$$

where $f_{V,l} = \pi R_l^2 / A_{\text{cell}}$ ($l$=1,2), $A_{\text{cell}} = |\mathbf{b}_1 \times \mathbf{b}_2|$ is the area of the unit cell and $J_1$ is the cylindrical Bessel function of the first kind and first order. Note that $p_{g,0} = g_b + \sum_{l=1,2} f_{V,l}\left(g_l - g_b\right)$.

In practice, the plane wave expansions are truncated in such a way that $|i_1|, |i_2| \leq n_{\max}$ and $|j_1|, |j_2| \leq n_{\max}$. Thus, each submatrix of $\mathbf{L_k}$ has dimension $(2n_{\max}+1)^2 \times (2n_{\max}+1)^2$ whereas $\mathbf{L_k}$ has dimensions $6(2n_{\max}+1)^2 \times 6(2n_{\max}+1)^2$. Note that each component of the state vector is expanded with a total of $(2n_{\max}+1)^2$ plane waves.

### II) Full spatial cutoff model

For the full spatial cutoff model the operator $\hat{L}_\mathbf{k}$ is determined by Eq. (S13). Proceeding as in the previous subsection, it is readily shown that it is represented by the following infinite matrix:

$$\mathbf{L_k} = \begin{pmatrix} 0 & 0 & \left[-(\mathbf{k}+\mathbf{G_I})\cdot\hat{\mathbf{y}}\delta_{\mathbf{I,J}}\right] & \left[-i\delta_{\mathbf{I,J}} \frac{1}{|\mathbf{k}+\mathbf{G_I}|^2/k_{\max}^2+1}\right] & 0 \\ 0 & 0 & \left[(\mathbf{k}+\mathbf{G_I})\cdot\hat{\mathbf{x}}\delta_{\mathbf{I,J}}\right] & 0 & \left[-i\delta_{\mathbf{I,J}} \frac{1}{|\mathbf{k}+\mathbf{G_I}|^2/k_{\max}^2+1}\right] \\ \left[-(\mathbf{k}+\mathbf{G_I})\cdot\hat{\mathbf{y}}\delta_{\mathbf{I,J}}\right] & \left[(\mathbf{k}+\mathbf{G_I})\cdot\hat{\mathbf{x}}\delta_{\mathbf{I,J}}\right] & 0 & 0 & 0 \\ i\frac{1}{c^2}\left[p_{\omega_p^2,\mathbf{I-J}}\right] & 0 & 0 & 0 & \left[-i\frac{\omega_c}{c}\delta_{\mathbf{I,J}}\right] \\ 0 & i\frac{1}{c^2}\left[p_{\omega_p^2,\mathbf{I-J}}\right] & 0 & \left[+i\frac{\omega_c}{c}\delta_{\mathbf{I,J}}\right] & 0 \end{pmatrix}$$

(S17)

## C. Continuous evolution of the hydrodynamic and full cutoff band diagrams to the local band diagram

In this Section, it is shown that the band diagrams of the hydrodynamic and full cutoff models evolve continuously to the band diagram of the local model, without closing the low-frequency gap. This property guarantees that there are no topological transitions in the low-frequency gap associated with the introduction of the cutoffs.

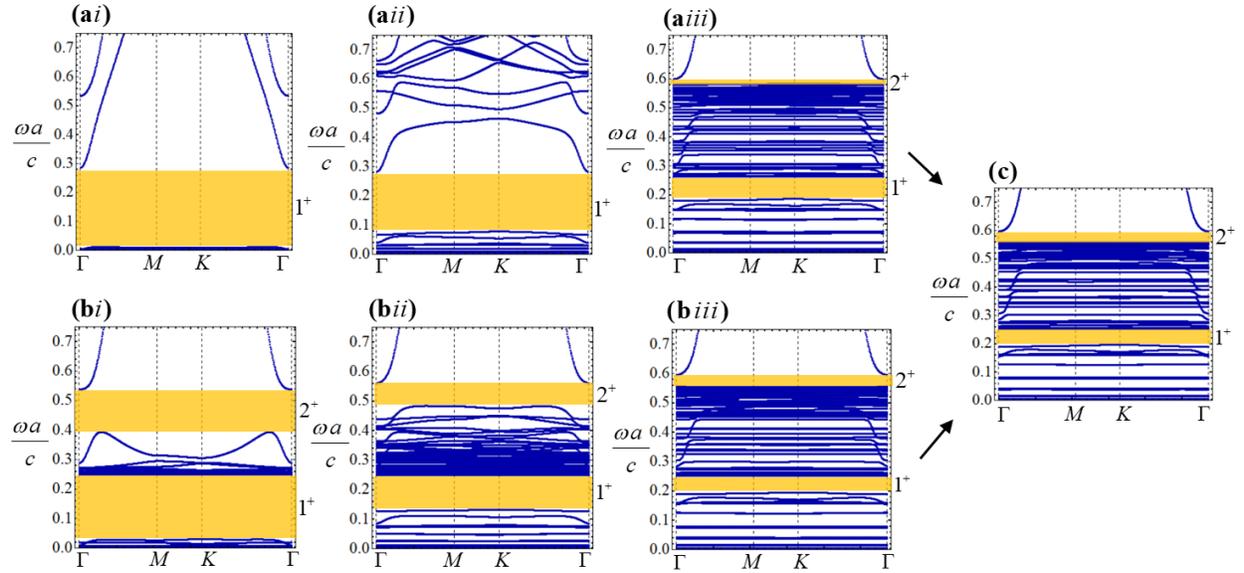

**Fig. S1** Evolution of the "nonlocal" band diagrams to the local case as $\beta$ and $1/k_{max}$ are varied continuously down to zero. **(a)** Band diagrams calculated with the hydrodynamic model with $\omega_c = 0.5\omega_p$ and **(ai)** $\beta = 0.5c$, **(aii)** $\beta = 0.1c$, and **(aiii)** $\beta = 0.01c$. **(b)** Band diagrams calculated with the full cutoff model with $\omega_c = 0.5\omega_p$ and **(bi)** $k_{max} = 2\omega_p/c$, **(bii)** $k_{max} = 10\omega_p/c$, and **(biii)** $k_{max} = 100\omega_p/c$. **(c)** Band diagram calculated with the local modal with $\omega_c = 0.5\omega_p$.

The first column of Fig. S1 depicts the same band diagrams as in the main text calculated with the hydrodynamic model ($\beta = 0.5$, panel $ai$) and with the full cutoff model ($k_{max} = 2\omega_p/c$, panel *b*i). The rest of the panels show how these band diagrams evolve continuously as $\beta \to 0$ and $k_{max} \to \infty$ to the corresponding local photonic crystal (panel *c*). As seen, the gaps of Fig. S1ai) and Fig. S1bi) remain open as $\beta$ and $1/k_{max}$ are varied continuously down to zero. Note that as $\beta$ approaches zero (for values on the order of $10^{-2}$) a second high-frequency gap is opened up in the hydrodynamic model.

## *D. Detailed numerical study of the convergence of the gap Chern number*

The gap Chern number is numerically calculated through the integral of Eq. (1) of the main text. The operator $\hat{L}_\mathbf{k}$ is replaced by a truncated version of the matrix given by Eq. (S15) (for the local and hydrodynamic models) or by Eq. (S17) (for the full cutoff model). Note that the derivatives $\partial_i \hat{L}_\mathbf{k}$ can be evaluated analytically. Each component of the state vector is expanded into $(2n_{max}+1)^2$ plane waves. In the numerical simulations the Brillouin zone is sampled with $N_1 \times N_2$ points, analogous to Ref. [S6]. Furthermore, in all the simulations the integral along the imaginary frequency axis ($\omega = \omega' + i\omega''$) is truncated at $-\omega''_{max} \leq \omega'' \leq \omega''_{max}$ with $\omega''_{max} a/c = 3$. The integral over $\omega$ is evaluated with the trapezoidal quadrature rule with $N_\omega$ sampling points. In all the simulations, we take $N_1 = N_2 \equiv N$.

Figure S2 shows the convergence analysis of the gap Chern number of the positive low-frequency gap (top row) and of the high-frequency gap (bottom row). As seen, the convergence of the gap Chern number is relatively fast in $n_{max}$, $N$ and $N_\omega$. As discussed in the main text, the high-frequency gap has a well defined topology ($\mathcal{C}^{2+}_{gap} = 1$), whereas the low-frequency gap has an ill-defined topology ($\mathcal{C}^{1+}_{gap} = -0.81$).

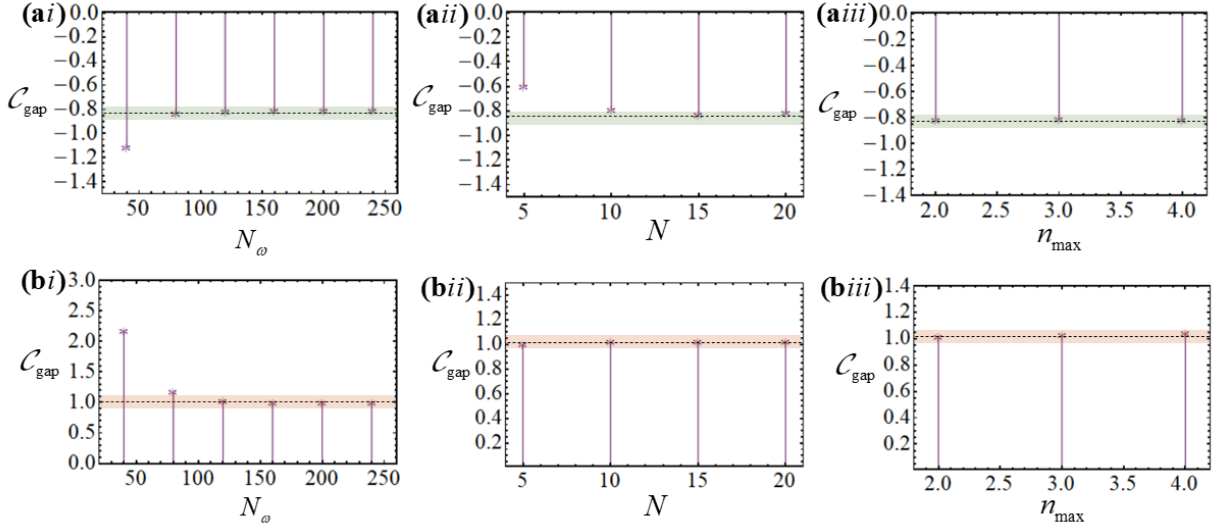

**Fig. S2** Convergence study of the numerically calculated gap Chern number $\mathcal{C}_{gap}$ for a photonic crystal described by the local model with the same parameters as in Fig. 2ai) as a function of (*i*) $N_\omega$, with $N = 20$ and $n_{max} = 3$. (*ii*) $N$, with $N_\omega = 120$ and $n_{max} = 3$. (*iii*) $n_{max}$, with $N = 20$ and $N_\omega = 120$. (**a**) Low-frequency gap Chern number (evaluated with $\omega_{gap} = 0.19\, c/a$). (**b**) High-frequency gap Chern number (evaluated with $\omega_{gap} = 0.73\, c/a$).

Similar convergence studies for the hydrodynamic model and full cutoff model are shown in Figs. S3 and S4. The hydrodynamic model yields a single bandgap defined by $0.01 < \omega a/c < 0.29$ with trivial topological charge $\mathcal{C}_{gap}^{1+} = 0$. As seen in Fig. S3ai) the convergence of the hydrodynamic model with $N_\omega$ is rather slow.

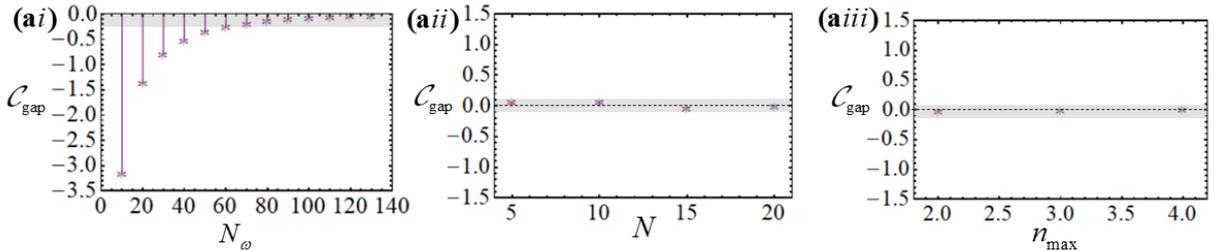

**Fig. S3** Convergence study of the numerically calculated gap Chern number $\mathcal{C}_{gap}$ for a photonic crystal described by the hydrodynamic model with the same parameters as in Fig. 2bi) as a function of (*i*) $N_\omega$, with $N = 20$ and $n_{max} = 3$.

(*ii*) $N$, with $N_\omega = 160$ and $n_{max} = 3$. (*iii*) $n_{max}$, with $N = 20$ and $N_\omega = 160$. The gap Chern number is evaluated using $\omega_{gap} = 0.15\,c/a$.

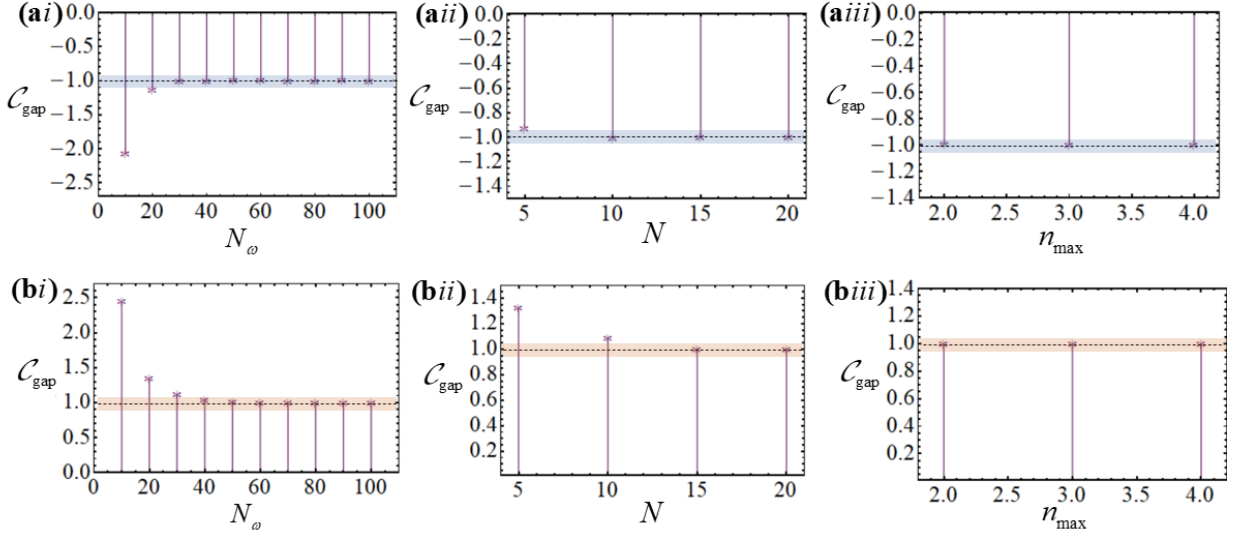

**Fig. S4** Convergence study of the numerically calculated gap Chern number $\mathcal{C}_{gap}$ for a photonic crystal described by the full cutoff model with the same parameters as in Fig. 2ci) as a function of (*i*) $N_\omega$, with $N = 20$ and $n_{max} = 3$. (*ii*) $N$, with $N_\omega = 120$ and $n_{max} = 3$. (*iii*) $n_{max}$, with $N = 20$ and $N_\omega = 120$. (**a**) Low-frequency gap Chern number (evaluated with $\omega_{gap} = 0.14\,c/a$). (**b**) High-frequency gap Chern number (evaluated with $\omega_{gap} = 0.5\,c/a$).

In contrast, the full cutoff provides a much faster convergence rate (Fig. S4). The two positive frequency gaps are defined by $0.4 < \omega a/c < 0.55$ and $0.03 < \omega a/c < 0.25$. The gap Chern number of the high (low) frequency gap converges to $\mathcal{C}_{gap}^{2+} = +1$ ($\mathcal{C}_{gap}^{1+} = -1$).

## E. Origin of the ill-defined Chern number

In this section, we explain the reason why an accumulation of branches at a single frequency may lead to an ill-defined Chern number.

Consider, for the purpose of illustration, the simplest case where it is possible to pick a gauge $\mathbf{Q}_{n\mathbf{k}}$ for the considered "band" that is globally defined and smooth in the entire Brillouin zone, with the exception of a single singular point $\mathbf{k}_S$. The band is formed by $N$ branches ($n = 1, 2, ..., N$); $N$ may be finite or infinite. The nontrivial case occurs when the branches cannot be disentangled, e.g., due to band crossings. Then, from topological band theory it follows that the Chern number of the band is $\delta \mathcal{C} = \frac{-1}{2\pi} \sum_{n=1}^{N} \oint_{\text{circle},\mathbf{k}_S} \mathcal{A}_{n\mathbf{k}} \cdot \mathbf{dl}$. Here, $\mathcal{A}_{n\mathbf{k}}$ is the Berry potential for the $n$-th branch and the line integral is over a circle of infinitesimal radius centered at the singular point. Usually one can guarantee that $\delta \mathcal{C}$ is an integer because it is possible to pick another basis $\mathbf{Q}'_{n\mathbf{k}} = \mathbf{Q}_{n\mathbf{k}} e^{i\theta_{n\mathbf{k}}}$ that is free from singularities in the neighborhood of $\mathbf{k}_S$ (see a detailed argument in [S8]). From topological band theory, the Berry potential determined by the new gauge is $\mathcal{A}'_{n\mathbf{k}} = \mathcal{A}_{n\mathbf{k}} - \nabla \theta_{n\mathbf{k}}$ and thereby:

$$\delta \mathcal{C} = \frac{-1}{2\pi} \sum_{n=1}^{N} \oint_{\text{circle},\mathbf{k}_S} \nabla \theta_{n\mathbf{k}} \cdot \mathbf{dl} = -\sum_{n=1}^{N} m_n. \tag{S18}$$

It was taken into account that the integral of $\mathcal{A}'_{n\mathbf{k}}$ over the circle of infinitesimal radius vanishes (because $\mathcal{A}'_{n\mathbf{k}}$ is free of singularities in the relevant neighborhood) and that the variation of $\theta_{n\mathbf{k}}$ over the circle must be an integer ($m_n$) multiple of $2\pi$ to ensure that both $\mathbf{Q}'_{n\mathbf{k}}$ and $\mathbf{Q}_{n\mathbf{k}}$ are smooth functions away from $\mathbf{k}_S$. When the considered band is formed by a finite number of branches ($N$ finite), the previous arguments imply that $\delta \mathcal{C}$ is an integer [S8]. In contrast, if the band is formed by an infinite number of branches, one sees that $\delta \mathcal{C}$ is given by an infinite sum of integers, which is generically divergent, for the reasons discussed in the main text. Thus, a band formed by an infinite number of branches usually leads to an ill-defined topology, explaining the results of the main text.

To conclude, we discuss why the full cutoff model regularizes the topology notwithstanding that it does not prevent the accumulation of bands at a single frequency. To begin with, we note that when there is an accumulation of flat-bands most of the eigenstates are associated with very localized resonances, and thereby their properties are determined by large values of $\mathbf{k}$. Importantly, as for the full cutoff model the permittivity of the bulk material satisfies $\bar{\varepsilon}_{\mathbf{k}\to\infty} \to \varepsilon_0 \mathbf{1}$, the eigenstates of most of the branches are expected to be very similar to those of a reciprocal photonic crystal (with trivial topology), and thereby the contribution of most terms in Eq. (S18) will vanish. Thus, the full cutoff guarantees that the sum (S18) is formed by a finite number of nonzero terms, justifying in this manner the topology regularization.

## Supplemental Material References: